\documentclass[aps,pra,a4paper,showpacs,twocolumn,floatfix]{revtex4}

\usepackage[latin1]{inputenc}
\usepackage{amsfonts}
\usepackage{amssymb}
\usepackage{amsmath}
\usepackage{calc}
\usepackage{graphicx}
\usepackage{epsfig}
\usepackage{bm}
\usepackage{ulem}
\usepackage{setspace}
\usepackage{subfigure}

\begin{document}

\title{Spin Wave Storage using Chirped Control Fields \\
in Atomic Frequency Comb based Quantum Memory}
\date{\today}
\author{Ji\v{r}\'i Min\'a\v{r}, Nicolas Sangouard, Mikael Afzelius, Hugues de Riedmatten, Nicolas Gisin}
\affiliation{Group of Applied Physics, University of Geneva, CH-1211 Geneva 4, Switzerland}

\begin{abstract} 
It has been shown that an inhomogeneously broadened optical transition shaped into an atomic frequency comb can store a large number of temporal modes of the electromagnetic field at the single photon level without the need to increase the optical depth of the storage material. The readout of light modes is made efficient thanks to the rephasing of the optical-wavelength coherence similarly to photon echo-type techniques and the re-emission time is given by the comb structure. For on-demand readout and long storage times, two control fields are used to transfer back and forth the optical coherence into a spin wave. Here, 
we present a detailed analysis of the spin wave storage based on chirped adiabatic control fields. In particular, we verify that chirped fields require significantly weaker intensities than $\pi$-pulses. The price to pay is a reduction of the multimode storage capacity that we quantify for realistic material parameters associated with solids doped with rare-earth-metal ions.
\end{abstract}

\maketitle

\section{Introduction}
\label{sec Introduction}

In future quantum networks, quantum information may be exchanged between the network nodes using trains of single-photons, i.e. photons prepared in temporally distinguishable modes, similarly to today's telecommunication networks. What we know with certitude is that temporal multiplexing greatly speeds-up the distribution rate of entanglement over long distances in quantum networks relying on quantum repeaters \cite{Simon07, Sangouard09}. This requires quantum memories enabling the storage of many temporal modes. However, the multimode capacity of most quantum memories based on atomic ensembles strongly depends on achievable absorption depths. Currently, only a few modes could efficiently be stored in available atomic ensembles using electromagnetically induced transparency, controlled and reversible inhomogeneous broadening or Raman-type memories \cite{Nunn08}. Time multiplexed quantum storage could thus be the bottleneck of future quantum networks.\\

However, tens of temporal modes at the single-photon level have been stored recently in a rare-earth-ions-doped solid with a rather limited optical depth \cite{Usmani10}. The key feature of this multimode storage has been presented in Ref.  \cite{Afzelius09} and consists in shaping the inhomogeneous broadening of an optical transition into an atomic frequency comb (AFC). When a photon enters the solid, the comb modes are excited. They dephase, then rephase at a time given by the comb structure leading to a photon-echo like reemission \cite{Kurnit64} in a well defined mode. For both on-demand read-out and long storage times, a pair of control pulses are used to transfer back and forth the optical coherence into a spin wave. This storage technique has already motivated a large number of proof-of-principle experiments \cite{Usmani10, DeRiedmatten08, Afzelius10, Chaneliere10, Amari10, Sabooni10, Chaneliere10b, Bonarota10}, because its multimode capacity only depends on the comb structure provided that the optical depth is high enough to store efficiently a single mode. However, the requirements on the control pulses have never been studied in detail. So far, it has been suggested that $\pi$-pulses could realize the desired population transfer over large frequency bandwidths. Here, we show that this necessitates high Rabi frequencies that are very challenging to achieve in rare-earth doped materials where the dipole moments are typically three orders of magnitude weaker than in the alkali gases usually used in quantum optics experiments. Since adiabatic chirped pulses are commonly used for the coherent control of rare-earth doped solids \cite{deSeze03, Crozatier04, deSeze05, Rippe05, Klein07} and are known to need weaker intensities than $\pi$-pulses \cite{Shore90}, we study the possibility to use them in AFC protocols. In particular, we study in detail the advantages of chirped pulses for light storage in solids and for limited Rabi frequencies, we highlight a tradeoff between efficient population transfer and high time multiplexing. \\

The paper is organized as follows. In the next section, we recall the principle of AFC memories. We then present a specific sequence of two chirped pulses which offers an efficient population transfer from the optical transition to the spin transition and preserves the collectivity that is at the heart of the read-out process. In section \ref{sec_Results}, we give explicit examples relevant to rare-earth doped solids to quantify the intensity gain and the multimode capacity losses when chirped pulses are used instead of conventional $\pi$-pulses. The last section contains our conclusions and an outlook toward future works. 

\section{AFC principle}
\begin{figure}
\includegraphics[width = 8.5 cm] {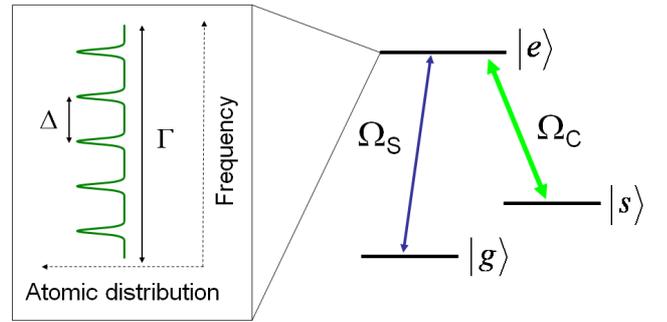}
\caption{(Color online) Principle of quantum memory for light  based on atomic frequency comb. The inhomogeneously broadened optical transition $|g\rangle$-$|e\rangle$ is shaped into an atomic frequency comb made of narrow peaks separated by $\Delta$ and spanning a large frequency range $\Gamma.$ When a resonant signal field (corresponding classically to the Rabi frequency $\Omega_s$) is absorbed, the comb modes are excited. They dephase, then rephase at time $2\pi/\Delta$ when the signal field is reemitted through a photon-echo type process. A pair of control pulses $\Omega_c$ on the $|e\rangle$-$|s\rangle$ transition allows for both on-demand read-out and long storage times.}
\label{fig1}
\end{figure}

Let us briefly recall the principle of light storage based on AFC. Consider an ensemble of atoms with an optical transition formed by two levels $|g\rangle$ and $|e\rangle,$ which is inhomogeneously broadened, see Fig. \ref{fig1}. The $|g\rangle$ - $|e\rangle$ transition is spectrally shaped such that the atomic distribution consists of a frequency comb made of narrow peaks with a characteristic width $\gamma,$ separated by $\Delta$ and spanning a large frequency range $\Gamma.$ When a signal photon enters the crystal, with a spectral profile centered on the frequency of the $|g\rangle$-$|e\rangle$ transition and with the bandwidth $\gamma_p$ satisfying $\Delta \ll \gamma_p \leq \Gamma ,$ it is completely absorbed provided that the atomic density is high enough. After the absorption, the photon is stored in a single atomic excitation delocalized over all the atoms, corresponding to the Dicke-type state \cite{Dicke54}
\begin{equation}
\label{state_vector}
\sum_{j=1}^{N} \rm{e}^{-i\Delta_j t} \rm{e}^{-ik_s z_j}|{g_1..e_j..g_N}\rangle,
\end{equation}
where $N$ is the number of atoms, $\Delta_j$ is the detuning of the j-th atom with respect to the laser frequency, $k_s$ is the wave number of the absorbed signal field and $z_j$ is the position of the atom $j$. 
(Note that in practice the amplitudes of the different terms may vary, depending e.g. on the laser profile.) For an atomic frequency comb with very sharp peaks, the detunings $\Delta_j$ are approximately multiples of $\Delta,$ i.e. $\Delta_j=m_j \Delta,$ $m_j \in \mathrm{Z}.$ After a time $2\pi/\Delta,$ the components $|{g_1..e_j..g_N}\rangle$ are in phase and lead to the echo-type re-emission of a photon \cite{Kurnit64} in a well defined mode which has the same wave number as the absorbed photon (forward re-emission). This is a remarkable feature of collective excitations: for an atomic ensemble that contains sufficiently many atoms, the emission in one mode completely dominates all other modes thanks to a collective interference. This allows for a very efficient retrieval of the stored photon. \\
The excellent multimode capacity of quantum memory based on AFC can be understood as follows. The shortest duration $\tau$ of one mode that can be efficiently stored in the memory is limited by the frequency bandwidth of the comb, i.e. $\tau \approx 12\pi/\Gamma$ (the pre-factor comes from the condition that the overlap of two adjacent temporal modes has to be negligible, see \cite{Afzelius09} for a detailed discussion). The total duration of the pulse train is limited by the echo time, i.e. the time it takes for the first mode to be reemitted, i.e. $2\pi/\Delta$. The number of modes that can be stored is given by the ratio between the pulse train duration and the duration of one mode, i.e. $\Gamma/(6 \Delta).$ This roughly corresponds to the number of peaks in the comb structure $N_{\rm{peak}}=\Gamma/\Delta$. In rare-earth-doped materials, the ratio between inhomogeneous broadening and homogeneous linewidth is inherently large making possible to prepare combs with hundreds of peaks and thus to store many modes without having to increase the absorption depth.\\

The technique based on AFC described so far only implements a memory with a predetermined storage time. For on-demand read-out of the stored electromagnetic field, the single collective excitation (\ref{state_vector}) has to be transfered from the optical-wavelength transition to a spin transition involving the state $|g\rangle$ and an ancillary state, say the state $|s\rangle.$ The spin transition do not have a comb structure so that the temporal evolution of the phase associated to each component of the collective state (\ref{state_vector}) is frozen. The resulting spin wave is thus described by 
\begin{equation}
\label{spin_state}
\sum_{j=1}^{N} \rm{e}^{-i\Delta_j (2\pi/\Delta-T_0)} \rm{e}^{-i(k_s-k_c) z_j}|{g_1..s_j..g_N}\rangle,
\end{equation} 
where $2\pi/\Delta-T_0$ is the delay between the control pulse and the signal photon
and $k_c$ is the wave number of the control field. To retrieve the stored photons, a counter-propagating control field transfers back the spin wave into the optical collective excitation
\begin{equation}
\label{state_emm}
\sum_{j=1}^{N} \rm{e}^{-i\Delta_j (t-T_{s})} \rm{e}^{-i(k-2k_c) z_j}|{g_1..e_j..g_N}\rangle,
\end{equation} 
leading to the desired reemission at time $t=2\pi/\Delta+T_s.$ On-demand read-out is thus achieved by controlling the delay $T_{\rm{s}}$ between the two control pulses. Note that the use of counter-propagating control fields forces the re-emitted light to propagate in the backward direction. This suppresses the potential re-absorption in the forward re-emission and leads, in principle, to quantum memories with unit efficiency \cite{Afzelius09, Sangouard07}. (This is not necessary if the atomic ensemble is embedded in an asymmetric cavity operated in the impedance matching condition where unit efficiency can be obtained without reversing the propagation direction of control fields \cite{Afzelius10b}.) Further note that the spin wave allows for long-lived storage since spin coherence lifetimes are generally longer than the optical coherence lifetimes.\\

So far, it has been considered that $\pi$-pulses,  i.e. resonant monochromatic fields shaped by an envelop with the characteristic duration $\tau_c$ and associated with the Rabi frequency $\Omega_c^{\max}=\max_{t} \Omega(t)$ such that $\Omega_c^{\max} \tau_c \sim \pi,$ could realize the desired population transfers. However, for an efficient conversion of the collective excitation from the optical transition to the spin transition, the field envelop has to be short in time such that its Fourier transform is larger than the comb bandwidth, $1/\tau_c \geq \Gamma.$ Together with the condition that the control pulse is a $\pi$-pulse, this leads to the requirement $\Omega_c^{\max} \geq \Gamma.$ High Rabi frequencies are thus required to insure efficient on-demand read-outs of AFC memories with $\pi$-pulses. As mentioned in the introduction, the use of control pulses with weaker intensities would greatly simplify the experiments. \\

Motivated by this last consideration, we analyze in what follows on-demand read-out of AFC-type memories with chirped adiabatic pulses, i.e. slowly varying fields with a time dependent frequency which satisfy $\Delta^{\max} \tau_c >1$ and $\Omega_c^{\max} \tau_c >1$ where $2\Delta^{\max}$ is the frequency range associated to the detunings (c.f. below for the exact adiabatic criteria). Intuitively, these chirped fields are attractive with respect to $\pi$-pulses. For chirped pulses with a Fourier transform dominated by the chirp, efficient transfer from the optical transition to the spin transition can be achieved provided $\Delta^{\max} \geq \Gamma.$ This allows one to use longer pulse durations than $\pi$-pulses, i.e. $\tau_c \geq 1/\Gamma$ so that the adiabatic requirements can be fulfilled for weaker Rabi frequencies $\Omega_c^{\max} \leq \Gamma.$ However, the total duration of the input pulses, including the train of pulses to store and one control pulse, is limited by the echo-time $2\pi/\Delta$. Consequently, there is thus a trade-off between weak Rabi frequencies and high multimode capacity.\\


\section{AFC protocol with chirped pulses}
\label{sec_Description}

In order to use adiabatic chirped pulses in memories based on AFC, we have first to check that they preserve the collectivity, i.e. the phases of each of the components involved in the Dicke state (\ref{state_vector}). This is a priori not obvious, since the adiabatic manipulation of atoms is usually accompanied with non-trivial phase evolutions including e.g. dynamical and geometrical phases \cite{Guerin03}.\\

\subsection{Notations} Let us recall the well known dynamics of a single two-level atom, with levels $s$ and $e,$ under a pulsed excitation. Consider that the system starts out in an arbitrary coherent superposition of states $s$ and $e.$ It is excited with the Rabi frequency $\Omega_c(t)=\Omega_c^{\max} g(t)$ $(g \in [0,1]$ $\forall t)$ by a control pulse propagating parallel to the z-axis with the wave number $k_c$ and corresponding to the field 
\begin{equation}
E_c(t) = \frac{\hbar}{\mu_{es}}\,\Omega_c(t)\,\rm{cos}\left[ \omega_c(t) t - 	k_c z + \phi_c \right].
\end{equation}
$\mu_{es}$ is the dipole moment of the $|e\rangle$-$|s\rangle$ transition. The frequency of the control pulse is made time dependent to describe the chirp. This leads to a time dependent detuning with respect to the atomic resonance $\omega_{se},$ $\Delta_j(t)=\omega_{se}^j-\dot{\omega_c}t-\omega_c=\Delta^{\max} f(t)+\Delta_j$ $(f \in [-1,1]$ $\forall t).$ Under the rotating wave approximation, the corresponding Hamiltonian in the basis $\{s, e\}$ is given by
\begin{equation}
\label{Ham}
H(t)=\hbar 
\begin{pmatrix}
	0 & -\frac{\Omega_c(t)}{2}\,e^{i\phi} \\
	-\frac{\Omega_c(t)}{2}\,e^{-i\phi} & \Delta_j(t) 
\end{pmatrix}.
\end{equation}
The dynamics is fully determined by the propagator $U(t,t_i)=\exp(-i\int_{t_i}^t ds H(s)).$ \\

\subsection{$\pi$-pulses} 
Let us recall briefly the case of a $\pi$-pulse. Under the assumptions that the laser frequency is time independent $(f(t)=0)$ and the Rabi frequency is larger than the detuning $\Omega_c^{\max} > \Delta_j$, the propagator takes the explicit form 
\begin{equation}
\label{prop_pi}
U_{A}(\tau_c,0)=
\begin{pmatrix} 
	\cos{\frac{A}{2}} & -i\sin{\frac{A}{2}} \\
	 -i\sin{\frac{A}{2}} &  \cos{\frac{A}{2}}
\end{pmatrix},
\end{equation}  
where $A=\int^{\tau_c}_{0} ds \Omega_c(s) \approx \Omega_c^{\max} \tau_c$ is the pulse area. Optimal population transfer is achieved  when $A=\pi$ which corresponds to a $\pi$-pulse.\\
Let us check that a pair of $\pi$-pulses can preserve the collectivity. At time $2\pi/\Delta-T_0$ after the photon absorption, a $\pi$-pulse transfers the atom j from $e$ to $s.$ The resulting state is obtained by applying $U_{A=\pi}(\tau_c,0)$ to the state (\ref{state_vector}) and corresponds to the state (\ref{spin_state}) up to an irrelevant global phase factor. A second $\pi$-pulse delayed by $T_s$ with respect to the first one, converts back the spin wave into an optical atomic excitation. The state is given by (\ref{state_emm}) and it leads to an echo-type re-emission at time $2\pi/\Delta+T_s.$ This confirms that $\pi$-pulses can be used to control the read-out time of AFC memories. However, it highlights the requirement that the Rabi frequencies have to be larger than all the detunings $\Omega_c^{\max} > \Delta_j$ $\forall j,$ i.e. larger than the overall comb spectrum. \\

\subsection{Chirped adiabatic pulses} 
Let us now focus on chirped adiabatic pulses. The Hamiltonian associated to the interaction between a chirped pulse and one atom now depends on time not only through the Rabi frequency but also through the detuning (see Eq. (\ref{Ham})). In the adiabatic regime where $\Omega_c(t)$ and $\Delta_j(t)$ are slowly varying in time, one can get an explicit expression for the propagator 
\begin{equation}
U(\tau_c,0)=\begin{pmatrix} 
	\rm{c_0 c_{\tau_c} } u_{-} + \rm{s_0 s_{\tau_c}}u_{+} & \rm{s_0 c_{\tau_c}}u_{-} - \rm{c_0 s_{\tau_c}}u_{+} \\
	\rm{c_0 s_{\tau_c}}u_{-} - \rm{s_0 c_{\tau_c}}u_{+} &  \rm{s_0 s_{\tau_c}}u_{-} + \rm{c_0 c_{\tau_c}}u_{+}
 \end{pmatrix}.
\end{equation}
 $\rm{c_0,}$ $\rm{s_0}$ ($\rm{c_{\tau_c}},$ $\rm{s_{\tau_c}}$) stand for $\cos\theta(t)$ and $\sin\theta(t)$ evaluated at initial $(t=0)$ (final $(t=\tau_c)$) time where $\theta$ is defined as $\tan2\theta(t)=-\Omega(t)/\Delta_j(t).$ $\rm{u\pm}=\rm{exp}\left( -i\int_{0}^{\tau_c} \frac{1}{2}(\Delta_j(s)\pm \sqrt{\Omega(s)^2+\Delta_j(s)^2}) \rm{d}s \right).$ If the detuning starts with a negative value and ends up to be positive, the propagator reduces to
\begin{equation}
U(\tau_c,0)=\begin{pmatrix} 
	0 & -u_{+} \\
	u_{-} & 0
 \end{pmatrix}
\end{equation} 
and describes a complete atomic transfer.\\
Let us check that the collectivity is preserved when the conversion of the optical excitation into a spin-wave excitation is realized with chirped pulses. We focus on the case where the two control pulses have the same chirp. The reason can be understood intuitively as follows : Consider that the first control pulse has a chirp going from $-\Delta^{\max}$ to $\Delta^{\max}$. An atom with a negative detuning will be transfered on the spin state sooner than an atom with a positive detuning. However, the latter is brought back to the excited state before the atom associated to a negative detuning if the chirp of the second control pulse ramps also from $-\Delta^{\max}$ to $\Delta^{\max}.$ If the two control fields have exactly the same chirp shape, atoms associated to different detunings will spend the same time in the excited state during the duration $2\pi/\Delta+T_S$ leading to a high collectivity provided that the dynamical and geometrical phases cancelled out. Following the scenario for the $\pi$-pulses, we start with the state (\ref{state_vector}). (Note that the characteristic duration of the chirped pulses are no more negligible with respect to the storage time in opposition to the $\pi$-pulses). After $2\pi/\Delta-T_0,$ two identical chirped pulses delayed by $T_s$ interact with the atomic ensemble leading to the state
\begin{eqnarray}
\nonumber
&&\sum_{j=1}^{N} e^{-i\Delta_j (t-T_s-\tau_c)} U(\tau_c,0)U(\tau_c,0)  |g_1..e_j..g_N\rangle\\
\nonumber
&=& -\sum_{j=1}^{N} e^{-i\Delta_j (t-T_s-\tau_c)} u_-^j u_+^j |g_1..e_j..g_N\rangle\\
\nonumber
&=& -\sum_{j=1}^{N} e^{-i\Delta_j (t-T_s-\tau_c)} e^{-i\int_{0}^{\tau_c} \Delta_j(s) \rm{d}s} |g_1..e_j..g_N\rangle\\
\nonumber
&=& -e^{-i \Delta^{\max} \int_{0}^{\tau_c} f(s) \rm{d}s} \sum_{j=1}^{N} e^{-i\Delta_j (t-T_s)} |g_1..e_j..g_N\rangle.
\end{eqnarray}
Omitting the global phase factor which is identical for each atom, one clearly sees that the collectivity is completely restored at time $t=2\pi/\Delta+T_s.$ Efficient read-out of AFC memories can thus be achieved with chirped pulses provided that they are adiabatic and that they have the same chirp.\\

\section{Numerical examples}
\label{sec_Results}

\subsection{Population transfer with chirped adiabatic pulses} We now consider concrete examples for atomic inversion with amplitude and frequency modulated control fields. We choose the model proposed in the 70's by Allen and Eberly \cite{Allen75} which involves control pulses with a hyperbolic secant temporal shape and a hyperbolic tangent chirp. This choice is motivated both by existing experiments in rare-earth doped solids \cite{Rippe05} and by the simplicity of the adiabatic criteria. 
Indeed, for
\begin{eqnarray}
\Omega_c(t)&=&\Omega_c^{\max} \rm{sech}(t/\tau_c), \\
\Delta_j(t)&=&\Delta^{\max} \tanh(t/\tau_c)+\Delta_j,
\end{eqnarray}
atomic inversion over the frequency range $\Gamma$ with finite efficiency $\eta$ is achieved provided \cite{Silver85}
\begin{subequations}
\label{ad_con}
\begin{align}
& 2\Delta^{\max} \sim \Gamma,  \label{ad_con_Gamma}\\ 
& \Delta^{\max} \tau_c \geqslant 2, \label{ad_con_Delta} \\
& \Omega_c^{\max} \sim \Delta^{\max} \sqrt{1-\left(\frac{\log{(1-\eta)}}{\pi\Delta^{\max}\tau_c}+1\right)^2}. \label{ad_con_Omega}
\end{align}
\end{subequations}
The first equation guarantees that the pulse frequencies overlap all the atomic spectrum. The second and third inequalities insure that the chirp and the Rabi frequency vary slowly in time. These equations offer a systematic method to realize an efficient population transfer over the set of atomic spectral components $\Gamma:$ \\
(i) First, one has to choose the chirp $\Delta^{\max}$ so that $2\Delta^{\max}=\Gamma.$ \\
(ii) Then, the pulse duration $\tau_c$ has to be at least equal to $4/\Gamma.$ Longer is the pulse duration, better is the selectivity of the population transfer. In the limit of very long control pulse duration $\tau_c \gg 4/\Gamma,$ the transfer follows a square function centered at $\Delta_j=0$ and with full width half maximum (FWHM) spectral bandwidth $\Gamma.$ We recall that the total duration of the input pulses, including the train of signal modes and the control pulse is limited by the echo time $2\pi/\Delta.$ More precisely, we will see below that $\tau_c$ is limited by $(2\pi/\Delta - 12\pi/\Gamma)/7.$\\
(iii) Finally, the transfer efficiency $\eta$ (for the resonant frequency $\Delta_j=0$) is given by the Eq. (\ref{ad_con_Omega}) and depends on the achievable Rabi frequency. This last equation shows that the use of long pulse durations decreases the required Rabi frequencies. The price to pay is a reduction of the number of modes that can be stored in the AFC memory since the total duration of input fields is limited by $2\pi/\Delta.$ Hence, there is a tradeoff on the control pulse duration $\tau_c$ : On-demand readouts with weak control fields favor $\tau_c \rightarrow (2\pi/\Delta - 12\pi/\Gamma)/7,$ whereas a high multimode capacity favors $\tau_c \rightarrow 4/\Gamma.$ \\ 

It is instructive to compare the population transfer with chirped pulses and with $\pi$-pulses. A spin transfer over $\Gamma$ with efficiency $\eta$ (minimum efficiency for each frequency) using $\pi$-pulses with hyperbolic secant temporal shape requires \cite{Silver85}
\begin{equation}
\label{pi_con}
\Omega_c^{\max} \sim \frac{\pi}{4} \frac{\Gamma }{\rm{arcsech} {\sqrt{\eta}}}.
\end{equation}
To transfer efficiently atoms spanning a given frequency domain $\Gamma$ using a $\pi$-pulse, a minimum Rabi frequency is required (see (\ref{pi_con})). This leads to a maximum temporal duration (since $\Omega_c^{\max} \tau_c$ is fixed) which can be compared to the minimum duration of a chirped pulse ($\tau_c \sim 4/\Gamma$). Furthermore, the minimum duration of a chirped pulse maximizes its Rabi frequency for fixed $\Gamma$ and $\eta.$ Let us take an example. To achieve a population inversion with $\eta=95\%$ efficiency, hyperbolic secant $\pi$-pulses are at least $8$ times shorter but needs to be 16 times more intense than chirped pulses corresponding to the Allen and Eberly model, independently of the frequency range $\Gamma.$ Note that this comparison is approximative since it does not take the transfer details into account (the formula (\ref{pi_con}) corresponds to the minimum transfer efficiency while (\ref{ad_con_Omega}) gives the transfer efficiency for the resonant frequency). For more accuracy, we performed numerical simulations of a classical light storage using the AFC technique.\\

\subsection{Numerical simulations} 
For numerical simulations, we choose an atomic comb inspired by recent experiments in praseodymium-doped solids \cite{Afzelius10}. We take an AFC composed of Gaussian peaks with a width $\gamma=2\pi \times 25$ kHz. The optical depth per peak is taken equal to $\alpha L=4.$ This leads to an optimal finesse close to $4$ \cite{Afzelius09}, corresponding to a peak separation of $\Delta=2\pi \times 100$ kHz. The comb is composed of $N_{\rm{peak}}=40$ peaks and thus spans a frequency range $\Gamma = 40 \times \Delta = 2\pi \times 4$ MHz. The storage efficiency is $\eta_{echo}$=25\% (forward readout without control fields, \cite{Afzelius09}). This comb allows to absorb and retrieve roughly $7$ Gaussian temporal modes with a duration $12\pi/\Delta=1.5\mu$s each and with negligible overlaps \cite{Simon07} within the time interval $2\pi/\Delta=10\mu$s. Furthermore, to control the multimode storage capacity, the control pulses are gated with a square function parametrized by its FWHM $T_{\rm{cut}}.$ For truncated hyperbolic secant pulses, choosing $T_{\rm{cut}} \sim 7 \tau_c$ is sufficient for the non-adiabatic corrections to be negligible. We chose either a cutoff $T_{\rm{cut}} \sim 7 \times 4/\Gamma =1.2 \mu$s corresponding to the minimal cutoff that fulfilled the adiabatic condition (\ref{ad_con_Delta}) and reducing the multimode capacity by 1 mode or $T_{\rm{cut}} \sim (2\pi/\Delta-12\pi/\Gamma)=8.8 \mu$s, the longer cutoff allowing to store a single mode.\\

\begin{figure}[h]
\begin{center}
\includegraphics[width = 9 cm] {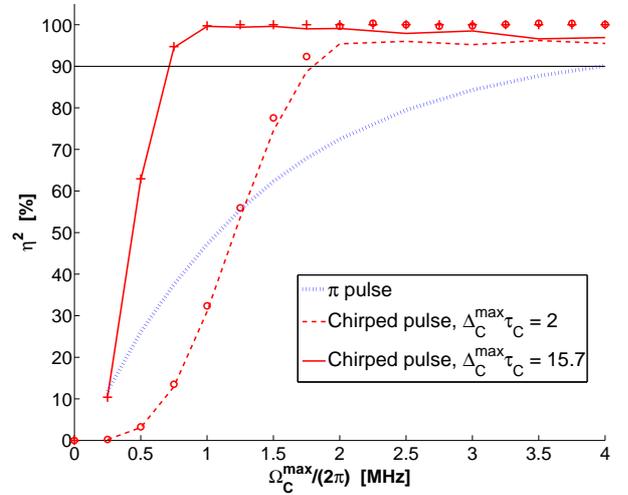}
\end{center}
\caption{(Color online) Transfer efficiency with $\pi$- (dotted blue line) and with chirped (red lines) pulses. The dashed line corresponds to chirped pulses of minimal temporal durations (satisfying the adiabatic criteria) for which the multimode capacity decreases of one mode. The full line is associated to chirped pulses of maximal temporal durations such that only one mode can be stored. The gain in term of the required intensity depends both on the desired efficiency and on the number of modes that can be sacrified. The circles and crosses are the values obtained using the equation (\ref{ad_con_Omega}). One can see a very good agreement with the numerical simulation.}
\label{fig2}
\end{figure}

The storage efficiency $\eta_{tot}$ is defined as the ratio of the echo intensity over the input intensity. It is the product of the absorption and reemission efficiency $\eta_{echo}$ with the transfer efficiency from the optical atomic excitation to the spin-wave squared $\eta^2$. Since one wants to characterize the transfer efficiency back and forth $\eta^2,$ we divide the overall storage efficiency $\eta_{tot}$ (with the spin wave transfer) by the echo efficiency without control fields $\eta_{echo}$. Figure \ref{fig2} shows $\eta^2$ for $\pi$-pulses and for chirped adiabatic pulses with $\Delta^{\max} \tau_c = 2$ ($T_{\rm{cut}}=1.2 \-\ \mu$s) and $\Delta^{\max} \tau_c = 15.7$ ($T_{\rm{cut}}=8.8\-\ \mu$s). One sees that both techniques can realize close to unit transfer efficiency. However, the adiabatic pulses require Rabi frequencies 2 to 5 times smaller than $\pi$-pulses for $\eta^2=0.9,$ leading to a gain between 4 and 25 in intensity. They are also longer in time and reduce the multimode capacity of the memory by at least 1 mode. Note that the number of lost modes can be made negligible with respect to the number of modes that can be stored by narrowing the individual peaks of the comb. For example, if $\gamma=2\pi \times 1$kHz can be achieved, the storage efficiency is optimized for a comb finesse of 4, corresponding to a peak separation of $\Delta=2\pi \times 4$kHz. In the time interval $2\pi/\Delta=250\mu$s, 160 modes can be stored with chirped pulses associated to $\Delta^{\max} \tau_c = 15.7$ instead of 166 for $\Delta^{\max} \tau_c = 2.$ \\

\begin{figure}[h]
\begin{center}
\includegraphics[width = 9 cm] {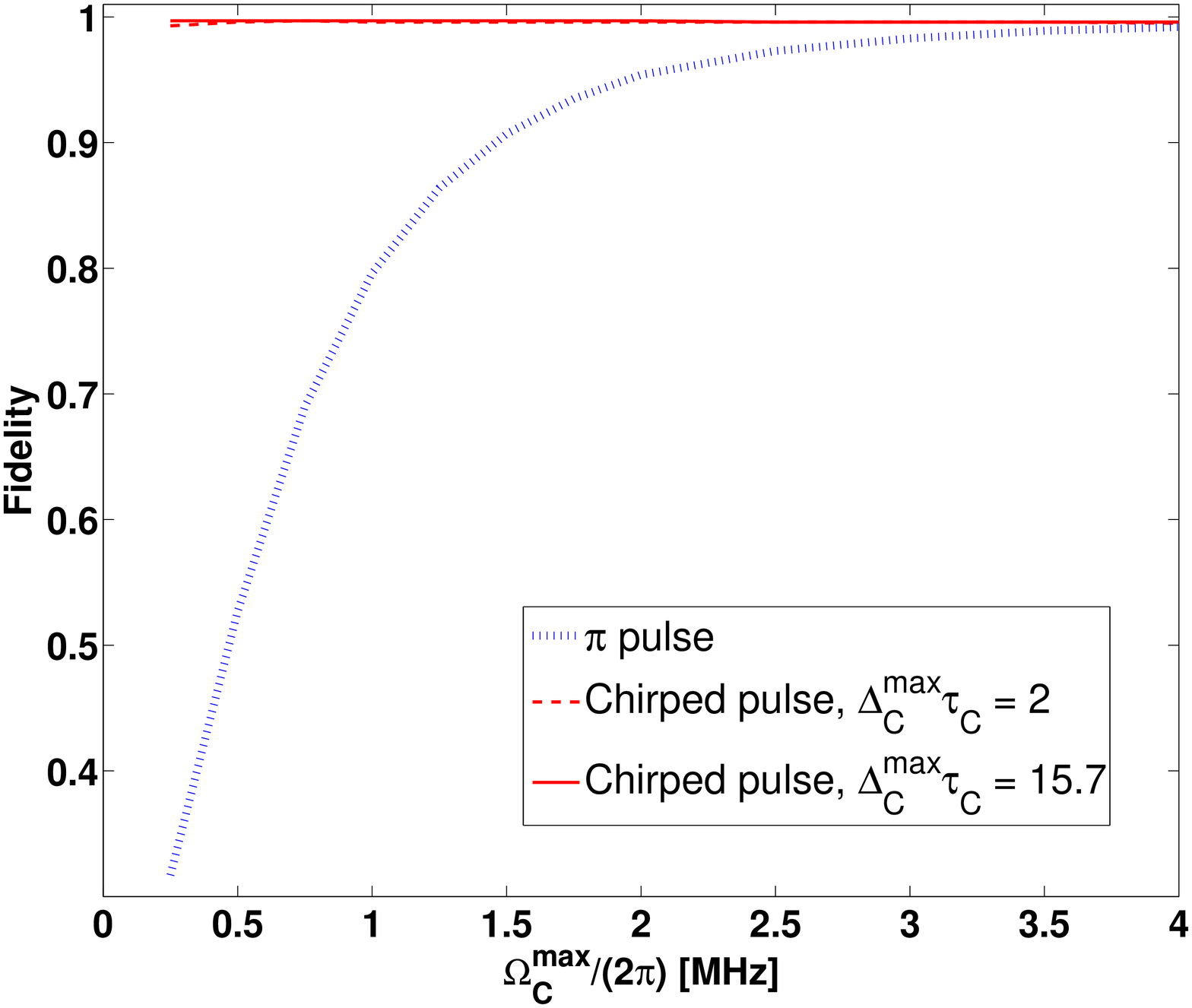}
\end{center}
\caption{(Color online) Overlap between the input signal and the retrieved field for $\pi$- (dotted blue line) and for chirped (red lines) pulses for different values of the Rabi frequency. As in Fig. \ref{fig2}, the dashed line corresponds to chirped pulses of minimal temporal durations (satisfying the adiabatic criteria) for which the multimode capacity decreases of one mode. The full line is associated to chirped pulses of maximal temporal durations such that only one mode can be stored.}
\label{fig3}
\end{figure}

Adiabatic methods are known to offer several advantages in practice over $\pi$-pulse based techniques. As detailed before, $\pi$-pulses are harder to implement due to laser power limitations. Furthermore, slowly varying chirped pulses allow for a population transfer with the same efficiency over a large frequency range leading to high storage fidelities. Figure \ref{fig3} shows the overlap between the input signal field and the output field as a function of the Rabi frequency. This permits one to see that only chirped pulses preserve the shape of the stored field for weak Rabi frequencies. This is essential in many applications involving interference effects where the retrieved photon needs to be indistinguishable from the input signal photon. Note also that the high selectivity of chirped pulses avoids unwanted transitions to nearby levels. Finally, let us recall that population inversion with $\pi$-pulses are not robust, e.g. with respect to Rabi frequency variations. For example, a small variation $\epsilon$ in the intensity of one of the control fields induces an error on the transfer efficiency $\eta^2$ of the order $\epsilon^2/2$ using $\pi$-pulses while it lets unchanged the transfer efficiency with chirped pulses provided that the adiabatic criteria are satisfied.

\section{Conclusion}
We showed how adiabatic methods can be used for on-demand readout of AFC memories. In particular, for weak Rabi frequencies, we highlighted a tradeoff between an efficient readout and a high multimode capacity. To get both an efficient readout and a multimode storage with achievable Rabi frequencies, the creation of a periodic narrow structure is crucial. In its ideal version, a AFC memory would be made of individual peaks as narrow as possible. Note that the width of individual peaks is limited ultimately by the homogeneous linewidth which can be of a few kHz in rare-earth doped solids. The peak separation is chosen to optimize the storage efficiency which can be very high provided that the absorption per peak is high enough. The multimode capacity is then mainly determined by the achievable Rabi frequency. For Rabi frequencies of the order of MHz, hundreds of modes could be efficiently stored in rare-earth doped solids. \\

For concreteness, consider a detailed example based on Europium-doped Y$_2$SiO$_5.$ Europium absorbs at 580 nm and have the appropriate lambda-energy structure. Since its homogeneous linewidth  
is very narrow ($2\pi \times 122$ Hz), one could realistically create a comb with $\gamma=2\pi \times 2$ kHz and $\Delta=2\pi \times20$ kHz over the spectral range $\Gamma=2\pi \times12$ MHz, i.e. a comb with $N_{\rm{peak}}=600$ peaks. The absorption coefficient is about 3-4 cm$^{-1}$ \cite{konz03} so that if an optical depth per peak of 40 is reached, the resulting echo efficiency is $\eta_{\rm{echo}}=90\%$ (backward readout). The storage of a single mode with a spin storage based on chirped pulses with the overall efficiency $\eta_{\rm{echo}} \eta^2 = 80\%$ requires $\Omega_c^{\max}=2\pi \times 0.5$ MHz. For $\Omega_c^{\max}= 2\pi \times 1$ MHz, 75 modes could be stored with the same efficiency. To fully profit from the multimode capacity, i.e. to store nearly 100 modes, $\Omega_c^{\max}=2\pi \times 5$ MHz is necessary.\\

In this paper, we focused on AFC memories assisted by chirped adiabatic pulses corresponding to the Allen and Eberly model to keep the experimental implementation simple. However, It would be of interest to consider alternative pulse shapes. For example, it has been shown in the frame of Nuclear Magnetic Resonance that the use of more sophisticated pulses requires Rabi frequencies up to two times smaller than the Allen and Eberly pulses \cite{Mitschang05}. Looking further ahead, it is an interesting question whether chirped pulses are optimal with respect to both the laser intensity and the multimode capacity. 

\section{Acknowledgments}
We acknowledge T. Chaneli\`ere, S. Gu\'erin, H.R. Jauslin, J.-L. Le Gou\"et, P. Sekatski and C. Simon for useful discussions and financial supports from the EU project QuReP and from the Swiss NCCR Quantum Photonics.

\end{document}